\documentclass[aip,reprint,cha,showpacs,amsmath,amssymb,superscriptaddress]{revtex4-1}
\usepackage{graphicx,SIunits,bm,array}
\usepackage[colorlinks=true,linkcolor=blue]{hyperref}

\begin{document}

\title{Ultra-low noise single-photon detector based on Si avalanche photodiode}

\author{Yong-Su \surname{Kim}}\email{yskim25@gmail.com}
\affiliation{Department of Physics, Pohang University of Science and Technology (POSTECH), Pohang, 790-784, Korea}

\author{Youn-Chang \surname{Jeong}}
\affiliation{Department of Physics, Pohang University of Science and Technology (POSTECH), Pohang, 790-784, Korea}

\author{Sebastien \surname{Sauge}}
\affiliation{Department of Microelectronics and Information Technology, Royal Institute of Technology (KTH), Electrum 229, SE-16440 Kista, Sweden}

\author{Vadim \surname{Makarov}}\email{makarov@vad1.com}
\affiliation{Department of Physics, Pohang University of Science and Technology (POSTECH), Pohang, 790-784, Korea}
\affiliation{Department of Electronics and Telecommunications, Norwegian University of Science and Technology (NTNU), NO-7491, Trondheim, Norway}
\affiliation{University Graduate Center, NO-2027 Kjeller, Norway}

\author{Yoon-Ho \surname{Kim}}\email{yoonho72@gmail.com}
\affiliation{Department of Physics, Pohang University of Science and Technology (POSTECH), Pohang, 790-784, Korea}

\date{September 22, 2011}

\begin{abstract}
We report operation and characterization of a lab-assembled single-photon detector based on commercial silicon avalanche photodiodes (PerkinElmer C30902SH, C30921SH). Dark count rate as low as $5\,\hertz$ was achieved by cooling the photodiodes down to $-80\,\celsius$. While afterpulsing increased as the photodiode temperature was decreased, total afterpulse probability did not become significant due to detector's relatively long deadtime in a passively-quenched scheme. We measured photon detection efficiency $>50\%$ at $806\,\nano\meter$.
\end{abstract}

%\pacs{85.60.Dw, 85.60.Gz}
% PACS, the Physics and Astronomy Classification Scheme.

%\keywords{Single photon avalanche diode, Low dark count}
%Use showkeys class option if keyword display desired

\maketitle

\section{Introduction}

Single-photon detectors (SPDs) are widely used for measuring extremely weak light levels, and they are an enabling technology for biophysics, astrophysics, metrology, quantum optics and quantum information experiments. Nowadays, there exist a number of competing technologies for SPDs, e.g.,\ photomultipliers,\cite{PMT1, PMT2} superconducting nanowire devices,\cite{SSPD1, SSPD2, SSPD3} visible light photon counters,\cite{VLPC2, VLPC1} and avalanche photodiodes (APDs).\cite{cova} When it comes to real-world applications, however, APDs are the most practical choice due to their good characteristics at relatively lower cost.

Silicon (Si) APDs are most often used in order to detect single photons in the visible and near-infrared wavelength range up to $\sim 1.1\,\micro\meter$.\cite{cova} Photon detection by Si APDs at longer wavelengths is also possible, enabled by efficient wavelength conversion to the above range.\cite{SFG1, SFG2, SFG3} 

Commercial SPDs based on Si APDs are available with dark count rates in tens to hundreds $\hertz$ range.\cite{SPCM, id, mpd, Tau-SPAD} Selected detector modules with extremely low dark count rates (below $2\,\hertz$) are also available, however they have a very small photosensitive area ($\sim$$20\,\micro\meter$ in diameter) and low photon detection efficiency in the near-infrared wavelength range ($\sim$$10\%$ at $780\,\nano\meter$).\cite{id} However, detectors with a very low dark count rate at a high ($>$50\%) photon detection efficiency can be important for a number of experiments and applications.\cite{ursin07, Lu07, Barz10}

In this paper, we report a lab-assembled SPD that exhibits dark count rates of only about $5\,\hertz$. Our SPD is based on PerkinElmer C30902SH or C30921SH APDs,\cite{pe} supplemented with our own electronics and cooling assembly. These APDs have a large input aperture (diameter of sensitive area $500\,\micro\meter$ or $250\,\micro\meter$) and high quantum efficiency in the 600--900$\,\nano\meter$ wavelength range. The paper is organized as follows: we describe the design of our lab-assembled SPDs, characterize them thoroughly, then compare their performance to that of commercial SPDs.

\section{Detector design}

We assembled several SPDs and tested nine APD samples: eight in window package (referred to as C30902SH-$n$, where $n$ is the sample number) and one in light-pipe package (C30921SH). Each APD was mounted on a thermoelectric cooler (TEC). We used 3-stage TEC (Thermonamic TEC3-71-31-17-03) to reach temperatures down to $-60\,\celsius$, and 4-stage TEC (Kryotherm TB-4-(83-18-4-1)-1.3) for lower temperatures. The TEC-APD assembly was mounted (using a thermally-conductive epoxy) inside an aluminum-alloy housing with an anti-reflection coated window. On the outside, the housing was mated with a low thermal resistance fan-cooled heatsink (Noctua NH-U12P) to keep the TEC hot plate close to the room temperature. In order to minimize heat flow to the cold plate, electrical connections to the APD and temperature sensor (miniature thermistor, RMT Ltd.\ TB04) were made with $50\,\micro\meter$ diameter platinum wires, allowing easy manual soldering. Furthermore, to reduce the heat flow via air convection, the TEC-APD assembly was tightly surrounded with cut-to-shape styrofoam (of ordinary type commonly found in, e.g.,\ shipping packaging material for consumer electronics). To prevent ice formation at the cold plate, the housing was sealed with rubber o-rings to keep it air-tight, and a small bag of strong desiccant was placed inside. For desiccant, we used calcium sulfate $\gamma$-anhydrite (Drierite) which worked well down to at least $-60\,\celsius$, and P$_{2}$O$_{5}$ (Sicapent) for lower temperatures at the cold plate.

\begin{figure}[t]
  \includegraphics[width=3.3in]{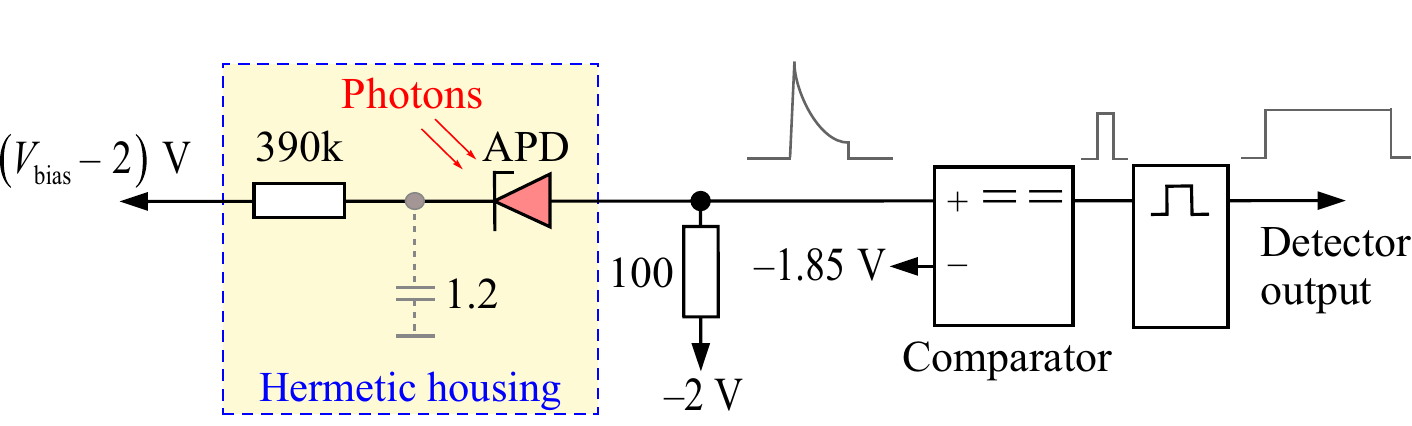}\\
  \caption{(Color online) Simplified circuit diagram of our passively-quenched SPD. The Si APD is biased above its breakdown voltage from a high-voltage source via the $390\,\kilo\ohm$ resistor. During an avalanche, the $1.2\,\pico\farad$ stray capacitance quickly discharges through the APD. The discharge current is converted into voltage at the $100\,\ohm$ resistor, and this voltage is sensed by a fast comparator (ECL differential receiver MC100EL16). When the voltage at the APD drops sufficiently close to the breakdown voltage, the avalanche quenches. Electrical signal shapes in the circuit are shown.}
  \label{quenching}
\end{figure}

We use a simple and robust passively-quenched detector circuit.\cite{cova,cova-quenching-circuits} It consists of a high-voltage power supply, $390\,\kilo\ohm$ bias resistor (placed close to the APD inside the hermetic housing), high-speed comparator sensing the onset of avalanche, and pulse width extender (Fig.~\ref{quenching}).\cite{circuit} The circuit works owing to the presence of a stray capacitance of about $1\,\pico\farad$ between the cathode of the APD and the ground. During the avalanche, this capacitance quickly discharges through the $100\,\ohm$ resistor, causing a short voltage pulse detected by a fast comparator. Detector electronics also contains a TEC controller that maintains a set temperature of the APD.\cite{circuit}

\section{Detector characteristics}

\subsection{APD breakdown voltage}

\begin{figure}[b]
  \includegraphics[width=3.3in]{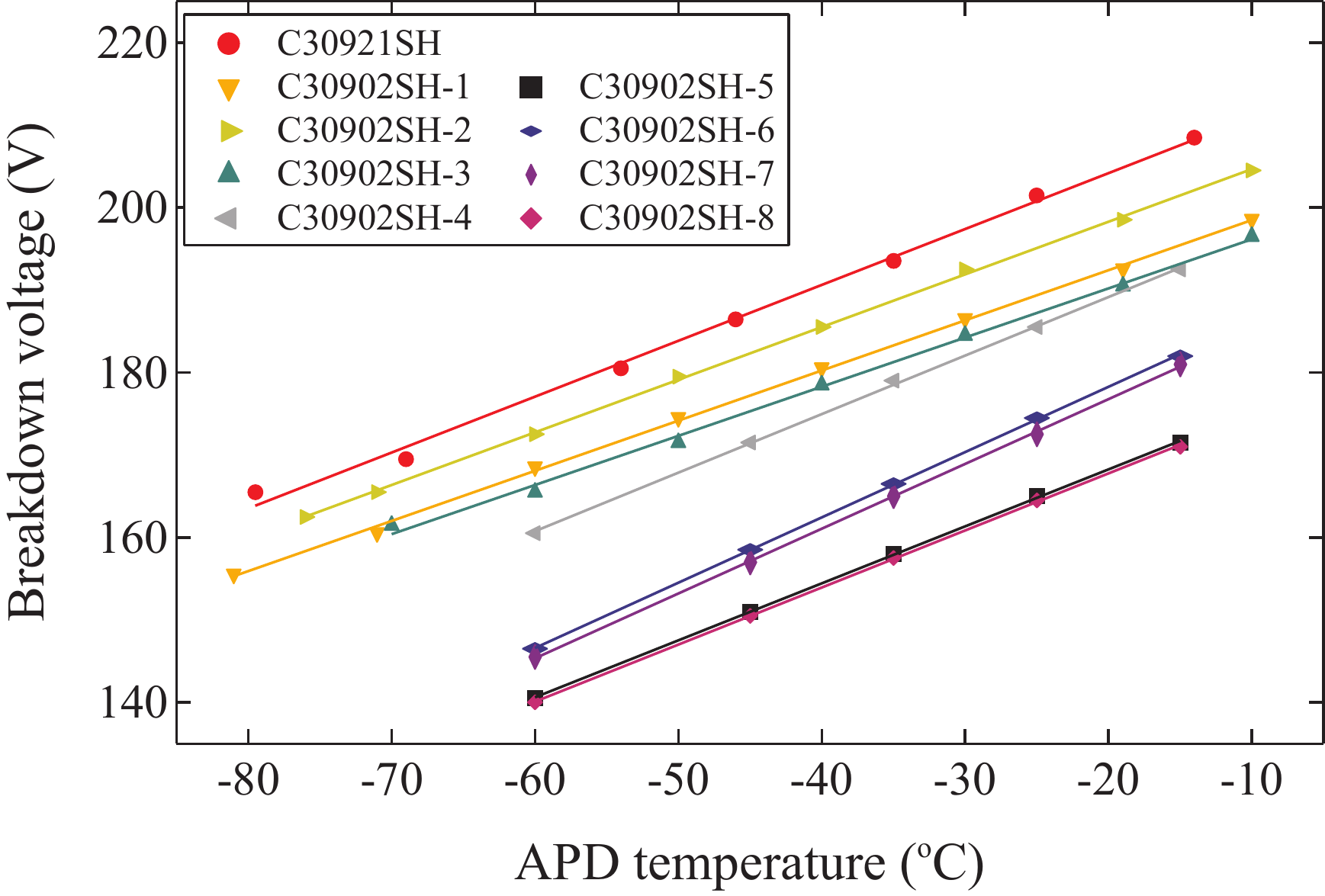}\\
  \caption{(Color online) Measured breakdown voltage as a function of APD temperature.}
  \label{V_break}
\end{figure}

We set the APD temperature $T_{\text{APD}}$ and bias voltage $V_{\text{bias}}$ with two trimpots. The tunability of $V_{\text{bias}}$ is required because the breakdown voltage $V_{\text{br}}$ increases with temperature. Figure~\ref{V_break} shows that $V_{\text{br}}$ depends on $T_{\text{APD}}$ linearly with a coefficient ranging from $0.60$ to $0.79\,\volt\per\celsius$ for different APD samples.

\subsection{Photon detection efficiency and dark count rate}

\begin{figure}[b]
  \includegraphics[width=3.3in]{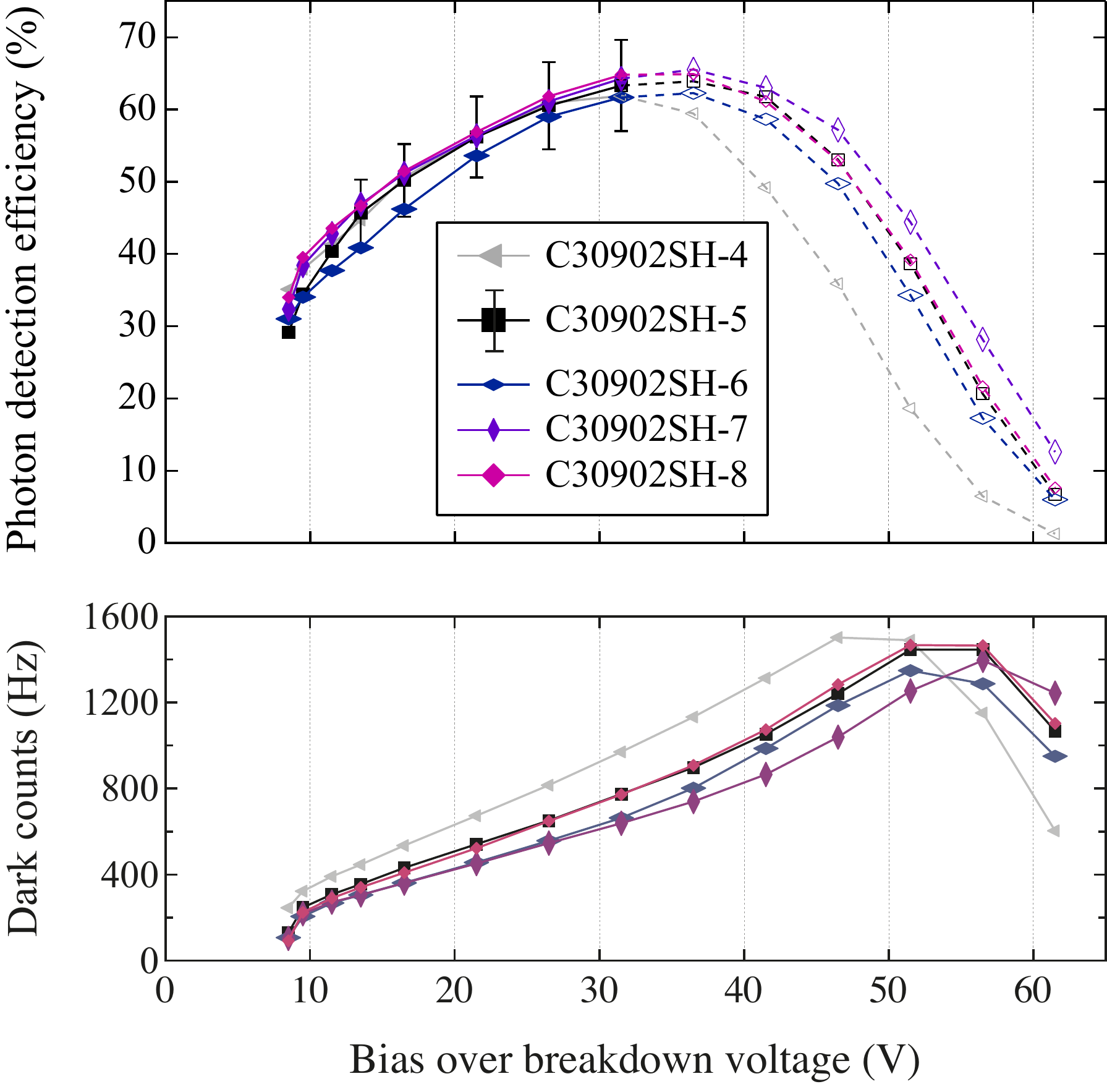}\\
  \caption{(Color online) (a) Calibrated photon detection efficiency (DE) and (b) dark count rate (DC) as a function of overbias voltage, at $-35$$\,\celsius$. DE measured at other temperatures in the $-80$ to $-10\,\celsius$ range is very similar. All measurements were performed at $806\,\nano\meter$, except for C30902SH-4 which was characterized at $834\,\nano\meter$. The error bars shown are $\pm 0.1$ of the measured DE value, which is the estimated maximum total error taking into account all calibration and measurement inaccuracies at $13.5\,\volt \le V_{\text{over}} \le 31.5\,\volt$. Data at $V_{\text{over}} < 8.5\,\volt$ was not measured because the avalanche pulse amplitude fell below the relatively high threshold of our comparator, set to avoid spurious counts from electrical interference. The dashed portions of the curves at $V_{\text{over}} > 30\,\volt$ are lower than the actual DE due to self-sustained avalanches, see Fig.~\ref{a-quenching} and discussion in the text. The drop of DC at $V_{\text{over}} > 40\,\volt$ is due to the same effect.}
  \label{QE}
\end{figure}

For characterization of our lab-assembled SPDs, we measured the photon detection efficiency (DE) and dark count rate (DC) as a function of overbias voltage $V_{\text{over}} = V_{\text{bias}} - V_{\text{br}}$. The detection efficiency $\text{DE}=(C-\text{DC})/N$ is obtained by measuring the count rate $C$ at a certain incoming photon rate $N=P\lambda/hc$, where $P$ is continuous-wave (c.w.)\ optical power focused in a spot in the middle of the APD photosensitive area, $\lambda$ is the laser wavelength, $h$ is the Planck constant and $c$ is the speed of light. The optical power illuminating the detector $P=P_l \times 10^{-Att/10}$ is obtained by measuring c.w.\ laser power $P_l$ with an optical power meter, then attenuating the light down to single-photon level with a set of calibrated attenuators (variable attenuator and neutral density filters) providing a cumulative attenuation $Att$ (dB).  

Figure~\ref{QE} shows DE and DC as a function of $V_{\text{over}}$ at $T_{\text{APD}}=-35\,\celsius$. We measured a maximum DE of $63 \pm 6\%$ at $31.5\,\volt$ overbias. A DE of $50 \pm 5\%$ at $17.5\,\volt$ overbias gives a possible trade-off between high DE and low DC. We remark that DE could be further increased slightly by decapsulating the APD package, whose entrance window is not anti-reflection coated.

\begin{figure}[t]
  \includegraphics[width=3.3in]{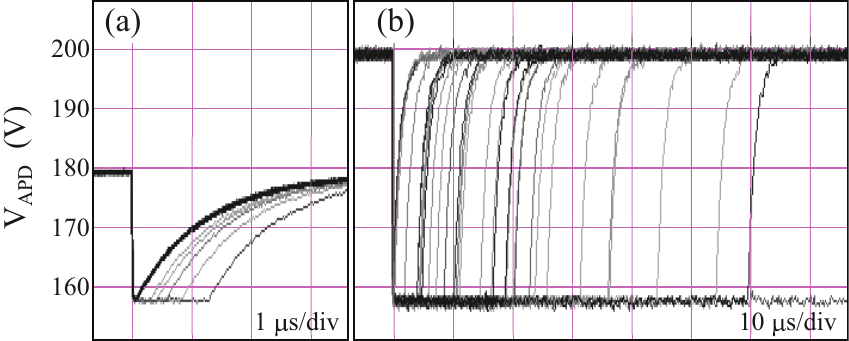}\\
  \caption{(Color online) Oscillograms of voltage across the APD showing avalanche self-quenching in the passively-quenched detector scheme. (a) At $V_{\text{over}}=21.5\,\volt$, most avalanches self-quench instantly, with just a few continuing for 1--2$\,\micro\second$ before self-quenching. (b) At $V_{\text{over}}=41.5\,\volt$, avalanches tend to self-sustain for tens of $\,\micro\second$, significantly extending the deadtime. At even higher overbias voltages the effect becomes runaway: the APD spends most of its time avalanching, insensitive to photons. The oscillograms were taken with C30902SH-8 at $-35\,\celsius$, $390\,\kilo\ohm$ bias resistor, and an oscilloscope with a time-limited display persistence.}
  \label{a-quenching}
\end{figure}

Hereafter, we investigate the drop of DE and DC at higher overbias voltages. In the passively-quenching circuit, the avalanche self-quenches quickly when the current through the APD drops below a certain value (latching current, $\sim 100\,\micro\ampere$).\cite{cova-quenching-circuits} The self-quenching happens because at a sufficiently low current, the number of carriers in the junction would randomly fluctuate to zero. In a self-sustaning avalanching condition, voltage at the APD drops to about $V_{\text{br}}$, thus the steady-state APD current equals $V_{\text{over}} / \text{value of the bias resistor}$. To achieve self-quenching, the bias resistor has to be sufficiently large and $V_{\text{over}}$ sufficiently low such that the steady-state current is lower than the latching current. We use $390\,\kilo\ohm$ bias resistor, which allows the avalanche to self-quench quickly at $V_{\text{over}} \lesssim 30\,\volt$. Figure~\ref{a-quenching} shows oscillograms of voltage across the APD when $V_{\text{over}}$ is $21.5\,\volt$ and $41.5\,\volt$. As can be seen, all avalanches self-quench within a few $\micro\second$ at the lower overbias voltage, whereas they self-sustain for a long time at the higher overbias, randomly extending the detector deadtime with a concomitant drop in measured DE and DC. Although using a larger bias resistor and $V_{\text{over}} > 30\,\volt$ could possibly result in a higher DE, we have noticed an uncontrolled rise of dark counts at $V_{\text{over}} > 80\,\volt$, and did not further investigate APD performance at those voltages. The remaining measurements in this paper are taken at $V_{\text{over}} = 17.5\,\volt$, and are thus not affected by the avalanche self-sustaining behavior.

\begin{figure}[t]
  \includegraphics[width=3.3in]{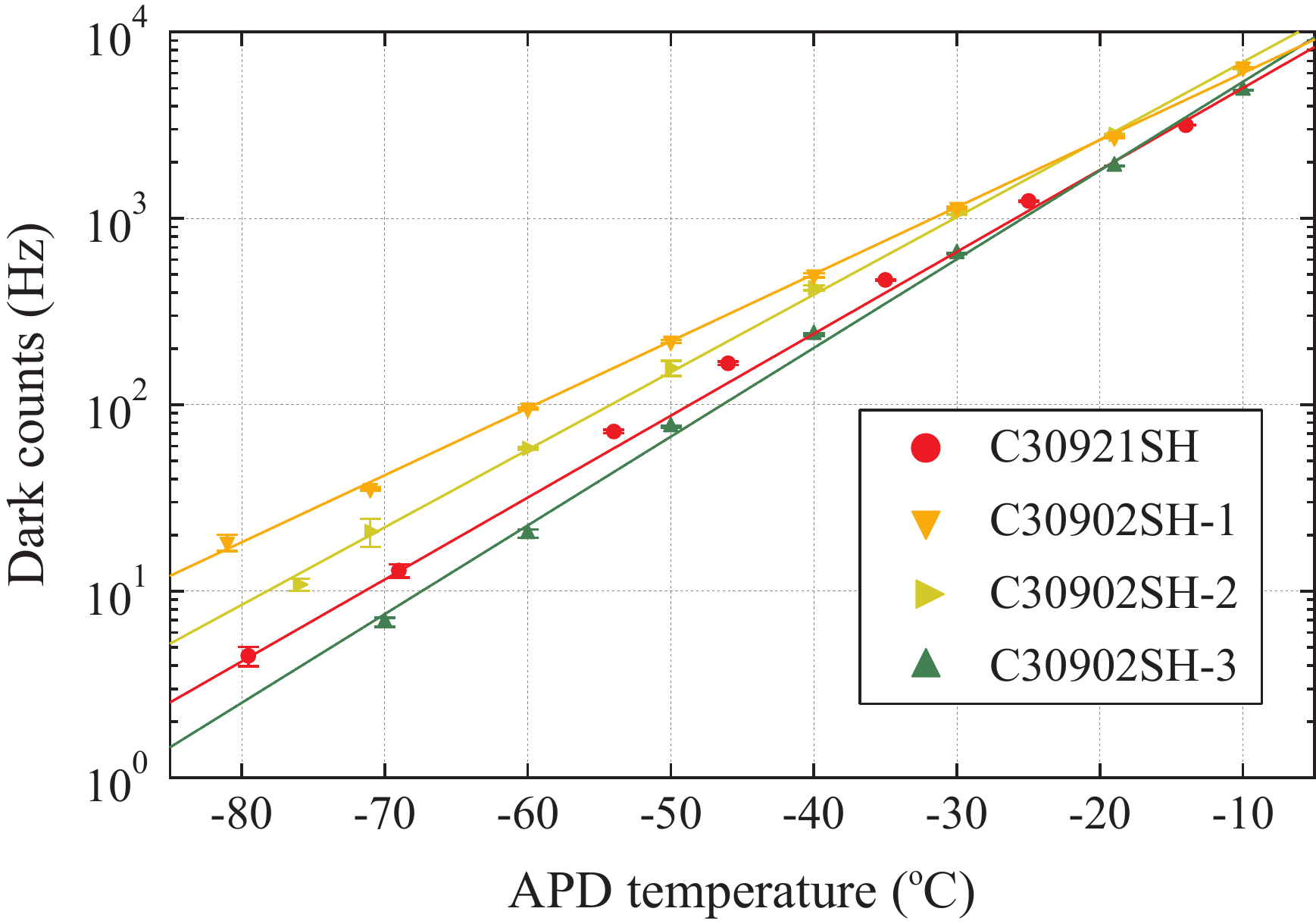}\\
  \caption{(Color online) Dark count rate (DC) as a function of APD temperature. Measurements were done at $V_{\text{over}}=17.5\,\volt$ for each temperature and sample. DC drops exponentially with decreasing temperature.}
  \label{dark_count}
\end{figure}

While DE in our measurement is insensitive of $T_{\text{APD}}$, DC increases exponentially with temperature (Fig.~\ref{dark_count}). On average, one can expect DC to halve for each $7.0\,\celsius$ decrease in temperature. At $-80\,\celsius$, we observed $\text{DC} < 5\,\hertz$ for the best sample, and $< 20\,\hertz$ for the worst one. We remark that temperatures down to $-30\,\celsius$ can easily be achieved with a single-stage TEC.

\subsection{Maximum count rate}

\begin{figure}[b]
  \includegraphics[width=3.3in]{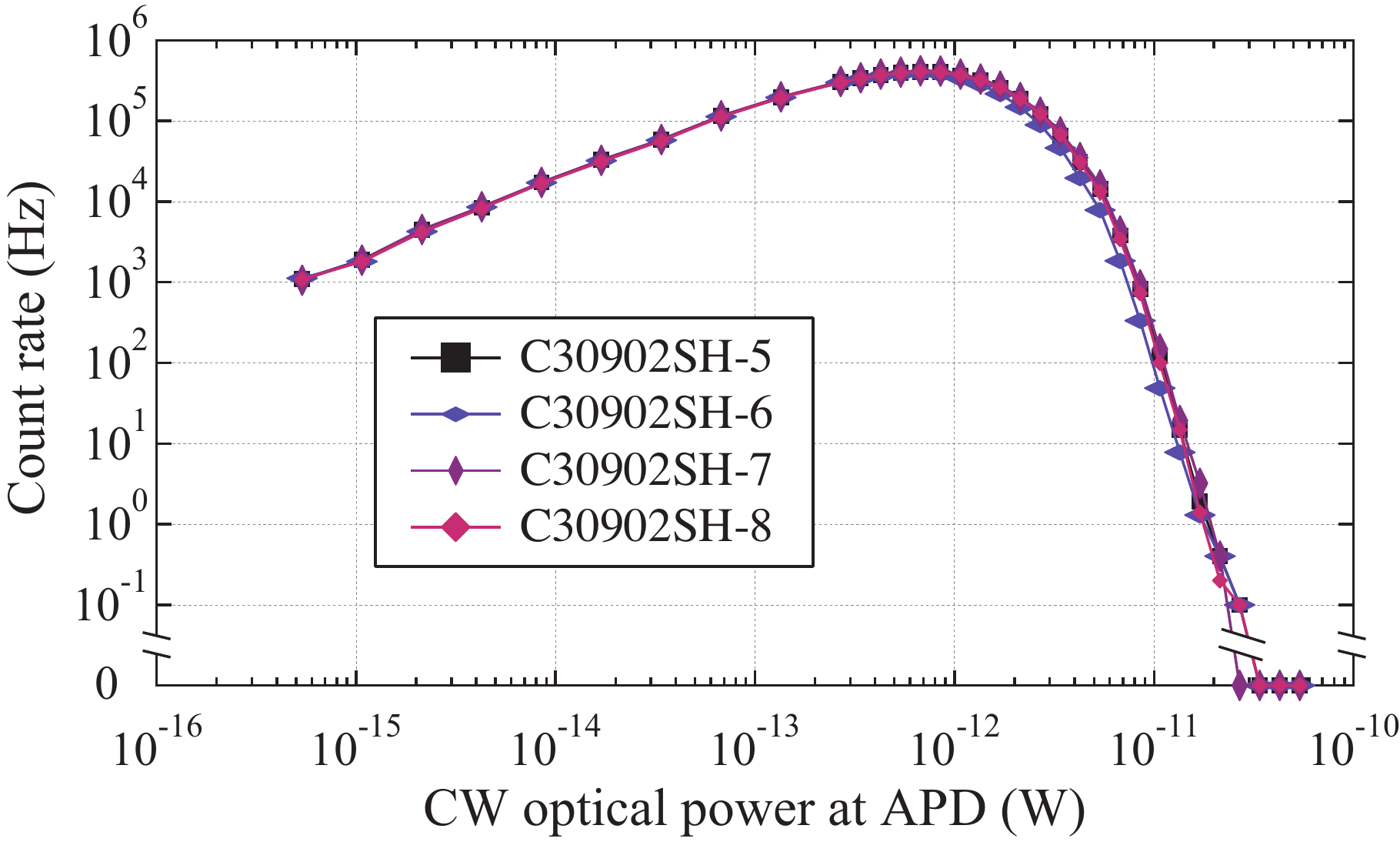}\\
  \caption{(Color online) Detector saturation curves as a function of calibrated optical power. Beyond the linear regime suitable for single-photon detection, the count rate saturates for optical power above $10^{-13}\,\watt$ and eventually drops to exactly zero under yet brighter illumination. Measurements were done at $-60\,\celsius$ and $V_{\text{over}}=17.5\,\volt$.}
  \label{saturation}
\end{figure}

Beyond the useful photon-counting range where the count rate grows linearly with optical illumination, passively-quenched SPDs exhibit saturation and blinding behavior.\cite{VM,Gerhardt2011} In case of our detectors, saturation starts at optical power above $100\,\femto\watt$ and the count rate peaks at $\sim 400\,\kilo\hertz$ (Fig.~\ref{saturation}). The SPDs become completely blind to single photons above $40\,\pico\watt$. These saturation and blinding curves are similar to other passively-quenched SPDs.\cite{VM}

\subsection{Afterpulsing}

Because of the relatively long deadtime of passively-quenched SPDs, afterpulsing is usually not as significant as with actively-quenched modules.\cite{cova,cova-quenching-circuits} However, since afterpulsing is known to increase as $T_{\text{APD}}$ decreases,\cite{afterpulsing} we have estimated afterpulsing probabilities of our lab-assembled SPDs at low temperatures. In order to measure them, we have measured the autocorrelation function $g^{2}(\tau)$ under complete darkness by following the time-correlated single-photon counting method.\cite{TCSPC} Usually for this purpose, $g^2(\tau)$ is measured under dim c.w.\cite{Becker05} or pulsed light.\cite{Cova91,Stipcevic10} Assuming $g^2(\tau)$ of photon counts is identical to that of dark counts, the latter is simpler to measure. Also it is more precise than under c.w.\ light, when afterpulsing probability is very small. Note that this is not the first example of using $g^2(\tau)$ of dark counts for quantifying afterpulsing probability.\cite{afterpulsing, Giudice09}

From $g^{2}(\tau)$, the afterpulsing probability $P_{T}^{ap}(\tau)$ is given by

\begin{equation}
P_{T}^{ap}(\tau)\approx[g^{2}(\tau)-1]\langle n_{T}\rangle,
\end{equation}
where $\tau$ is a delay time, and $\langle n_{T}\rangle\ll1$ is the average dark count probability per sample time $T$.\cite{AO_Rarity_2000} Since $\langle n_{T}\rangle$ is a constant over the delay time $\tau$, the afterpulsing probability $P_{T}^{ap}(\tau)$ can be determined by $g^{2}(\tau)$ alone. In the limit when $g^{2}=1$ (no correlation between counts), we get $P_{T}^{ap}=0$ (no afterpulsing). $g^{2}>1$ (resp. $g^2<1$) is to be understood as an increased (resp. decreased) probability of recording a correlated count, causing $P_{T}^{ap}>0$ (resp. $P_{T}^{ap}<0$, which is to be interpreted as detector deadtime and zero afterpulsing).

\begin{figure}[t]
  \includegraphics[width=3.3in]{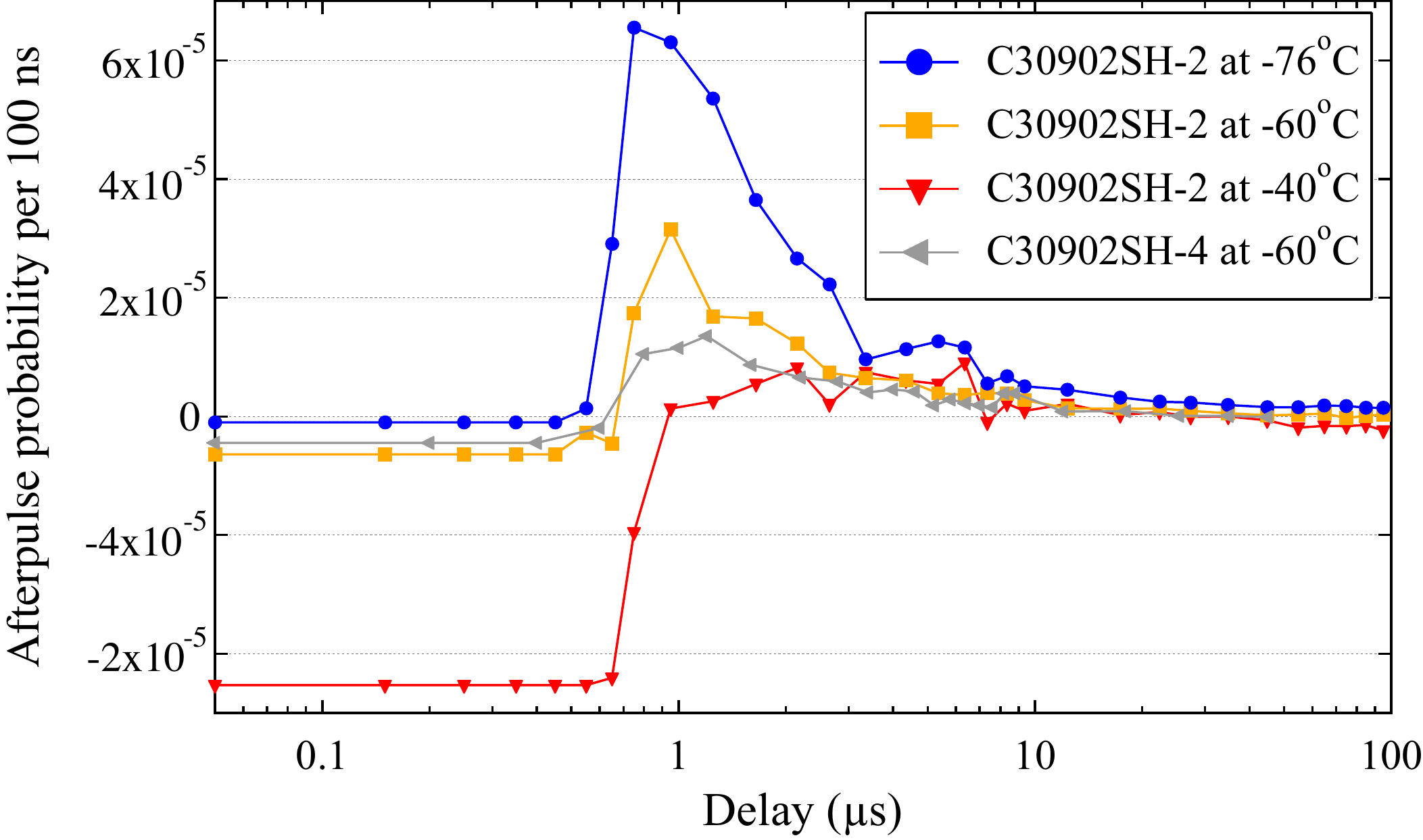}\\
  \caption{(Color online) Afterpulsing probability per sample time $T=100\,\nano\second$ as a function of time delay $\tau$. Measurements were done at $V_{\text{over}}=17.5\,\volt$. As $T_{\text{APD}}$ decreases, the afterpulsing probability increases. However, thanks to long deadtimes, the worst-case total afterpulsing probability over the first $10\,\micro\second$ is only 0.15\% at the lowest temperature. Note that the afterpulsing characteristics of the two different APDs at the same temperature are similar.}
  \label{afterpulse}
\end{figure}

Figure~\ref{afterpulse} shows afterpulsing probabilities for two APD samples. For the first few hundreds of $\nano\second$, $P_{T}^{ap}(\tau)$ is negative because $g^{2}(\tau)=0$ over the detector deadtime. From Eq.~(1), the magnitude of  $P_{T}^{ap}(\tau)$ is set by the dark count probability $\langle{n_T}\rangle$ for $g^{2}(\tau)=0$. This is why there are different non-zero levels for different samples and $T_{\text{APD}}$ during the deadtime.

\begin{table*}[t]
\renewcommand{\tabcolsep}{8pt}
\renewcommand{\arraystretch}{1.3}
\centering
\caption{Comparison of commercial Si-based SPDs and our lab-assembled SPD.}
\begin{ruledtabular}
\begin{tabular}[t]{m{48mm}|>{\centering}m{17mm}>{\centering}m{19mm}>{\centering}m{17mm}>{\centering}m{17mm}c}
Detector model & SPCM-AQRH\cite{SPCM} & id100-20\cite{id} & PDM series\cite{mpd} & Tau-SPAD\cite{Tau-SPAD} & our SPD\\
\hline
Peak detection efficiency &65\% ($650\,\nano\meter$)& 35\% ($500\,\nano\meter$) & 49\% ($550\,\nano\meter$)  & 75\% ($670\,\nano\meter$)\footnotemark[1] &\\
Detection efficiency at $806\,\nano\meter$ & 55\% & 7\% & 14\% & 59\%\footnotemark[1] & 50\%\footnotemark[2]\\
Dark count rate ($\hertz$) & typ.\ 250\footnotemark[3] & $< 60$\footnotemark[4] & $< 25$\footnotemark[5] & $< 20$ & typ.\ 10\\
Sensitive area diameter ($\micro\meter$) & 180 & 20 & 20\footnotemark[5] & 150 & 500\\
Dark count rate per unit area ($\hertz\per\micro\meter^{2}$) & $10^{-2}$ & $< 2\times10^{-1}$ & $< 8\times10^{-2}$ & $10^{-3}$ & $5\times10^{-5}$\\
Afterpulsing probability & 0.5\% & $<3$\% & $<3$\% & $<1$\%\footnotemark[1] & 0.15\%\\
Saturation count rate ($\mega\hertz$) & 25 & 20 & $> 10$\footnotemark[6] & $> 10$\footnotemark[7] & 0.4\\
\end{tabular}
\end{ruledtabular}
\footnotetext[1]{In the center of the photosensitive area.}
\footnotetext[2]{Although a higher DE can be achieved, we report the value measured at $17.5\,\volt$ overbias voltage, which we have chosen as a trade-off setting between high DE and low DC.}
\footnotetext[3]{$25\,\hertz$ unit is available although it is about twice as expensive ($\sim\$12,000$).}
\footnotetext[4]{Less than $2\,\hertz$ unit is available on a special order.}
\footnotetext[5]{Specification for PD2CTA model.}
\footnotetext[6]{Not specified in the datasheet; we infer it from the specified deadtime of $\sim 77\,\nano\second$.}
\footnotetext[7]{Not specified in the datasheet; we infer it from the specified deadtime of $< 60\,\nano\second$.}
\label{table:det-comparison}
\end{table*}

For $T_{\text{APD}} >-40\,\celsius$, there is hardly any measurable afterpulsing. As $T_{\text{APD}}$ decreases, the afterpulsing probability starts to rise. However, even at $-80\,\celsius$ the total afterpulsing probability within the first $10\,\micro\second$ is only 0.15\%. This low probability is due to detector's relatively long deadtime. Since the probability of afterpulsing decays exponentially from the time of avalanche,\cite{cova, afterpulsing} most of the trapped carriers are released during the long detector deadtime without causing a count. Note that cooling the APD to the lowest temperature will not necessarily provide the lowest spurious count rate, because in the first few $\micro\second$ the afterpulsing probability would dominate. The optimum temperature to achieve the lowest noise depends on the photon count rate in a given application, with lower temperature desirable at lower count rates. However, with this deadtime it never makes sense to set temperature above $-40\,\celsius$.

\section{Comparison to commercial detectors}

Before we conclude, let us compare characteristics of our lab-assembled SPD to other widely-used commercial Si-based SPDs. Table~\ref{table:det-comparison} shows primary characteristics of various SPDs. Our lab-assembled detector has one of the lowest dark count rates, the largest sensitive area, and a fairly high photon detection efficiency. In particular, our SPD has several orders of magnitude lower dark count rate per unit area than the other detectors. The maximum count rate of our lab-assembled SPD, however, is much smaller than the other SPDs, because we use the passive quenching scheme. One should note, however, that many quantum optics and quantum information experiments do not need such a high count rate.\cite{kim10} Thus, our lab-assembled SPD provides a valuable alternative. We remark that it costs under \$1000 in components.

\section{Conclusion}

In conclusion, we have built and characterized a lab-assembled SPD based on passively-quenched Si APDs, which exhibit a very low dark count rate ($<5\,\hertz$ at $-80\,\celsius$ for the best sample) and which simultaneously have both a high photon detection efficiency ($> 50\%$ at $806\,\nano\meter$) and a large photosensitive area ($500\,\micro\meter$ diameter). In the passively-quenched scheme, afterpulsing probability remains low thanks to detector's longer deadtime. In view of the relatively low photon detection efficiency (about 10\% at $780\,\nano\meter$) and small sensitive area ($20\,\micro\meter$ diameter) of low dark count commercial detectors,\cite{id} our lab-assembled SPD provides an alternative choice for quantum optics and quantum information experiments, as long as a high count rate is not required.

\section*{Acknowledgments}

This work was supported by the National Research Foundation of Korea (grant nos.\ 2009-0070668 and 2009-0084473) and the Research Council of Norway (grant no.\ 180439/V30). Y.-S.K.\ acknowledges support from the Korea Research Foundation (KRF-2007-511-C00004). V.M.\ acknowledges support from the BK21 program. S.S.\ acknowledges support from the ECOC'2004 foundation and the Research Council of Sweden (VR, grant no.\ 621-2007-4647).


\begin{thebibliography}{99}
\bibitem{circuit} The electronics has originally been designed by C.~Kurtsiefer and his coworkers, and modified by us.
\end{thebibliography}
\end{document}